\documentclass[fleqn,usenatbib]{mnras}

\usepackage{newtxtext,newtxmath}
\usepackage{mathrsfs}

\usepackage[T1]{fontenc}
\usepackage{ae,aecompl}
\usepackage{booktabs}

\DeclareRobustCommand{\VAN}[3]{#2}
\let\VANthebibliography\thebibliography
\def\thebibliography{\DeclareRobustCommand{\VAN}[3]{##3}\VANthebibliography}

\usepackage{graphicx}	
\usepackage{amsmath}	

\usepackage{amssymb}	

\title{Constraining violations of the Weak Equivalence Principle Using CHIME FRBs}

\author[Kaustubha Sen et al.]{
Kaustubha Sen$^{1,2}$\thanks{E-mail:ksen1993@gapp.nthu.edu.tw},
Tetsuya Hashimoto$^{3,2,1}$,
Tomotsugu Goto$^{1,2}$,
Seong Jin Kim$^{1,2}$,
\and
Bo Han Chen$^{1,2}$,
Daryl Joe D. Santos$^{2}$,
Simon C. C. Ho$^{1,2}$,
Alvina Y. L. On$^{2,5}$,
\and
Ting-Yi Lu$^{2}$,
and Tiger Y.-Y. Hsiao$^{2}$
\\
$^{1}$Department of Physics, National Tsing Hua University, No. 101, Section 2, Kuang-Fu Road, Hsinchu City 30013, Taiwan\\
$^{2}$Institute of Astronomy, National Tsing Hua University, No. 101, Section 2, Kuang-Fu Road, Hsinchu City 30013, Taiwan\\
$^{3}$Department of Physics, National Chung Hsing University, 145 Xingda Rd., South Dist.,
Taichung City, 402, Taiwan\\
$^{4}$Centre for Informatics and Computation in Astronomy (CICA), National Tsing Hua University, 101, Section 2. Kuang-Fu Road, Hsinchu, 30013, Taiwan\\
$^{5}$Mullard Space Science Laboratory, University College London, Holmbury St Mary, Surrey RH5 6NT, UK
}

\date{Accepted 2021 November 8. Received 2021 November 8; in original form 2021 October 13}

\pubyear{2021}

\begin{document}
\label{firstpage}
\pagerange{\pageref{firstpage}--\pageref{lastpage}}
\maketitle
\begin{abstract}
Einstein’s General Relativity (GR) is the basis of modern astronomy and astrophysics. Testing the validity of basic assumptions of GR is important. In this work, we test a possible violation of the Weak Equivalence Principle  (WEP), i.e.,  there might be a time-lag between photons of different frequencies caused by the effect of gravitational fields if the speeds of photons are slightly different at different frequencies. We use Fast Radio Bursts (FRBs) , which are astronomical transients with millisecond timescales detected in the radio frequency range. Being at cosmological distances, accumulated time delay of FRBs can be caused by the plasma in between an FRB source and an observer, and by gravitational fields in the path of the signal. We segregate the delay due to dispersion and gravitational field using the post-Newtonian formalism (PPN) parameter $\Delta \gamma$, which defines the space-curvature due to gravity by a unit test mass. We did not detect any time-delay from FRBs but obtained tight constraints on the upper limit of $\Delta \gamma$. For FRB20181117C with $z = 1.83 \pm 0.28$ and $\nu_{obs}$ =  $676.5\,{\rm MHz}$, the best possible constraint is obtained at log($\Delta \gamma$) = $-21.58 ^{+0.10}_{-0.12}$ and log($\Delta \gamma$/$r_{\rm E}$) = $-21.75 ^{+0.10}_{-0.14}$, respectively, where $r_{\rm E}$ is the energy ratio of two photons of the same FRB signal. This constraint is about one order of magnitude better than the previous constraint obtained with FRBs, and five orders tighter than any constraint obtained using other cosmological sources.  

\end{abstract}




\section{Introduction}
\label{section:Introduction}

Fast Radio Bursts (FRBs) are astronomical transient signals or millisecond time scales occurring at cosmological distances~\citep[]{Thornton_2013, Petroff_2016}. Since the first discovery in 2007 from the Parkes Data taken by~\citet[]{2008AIPC..983..584M}, FRBs have been an interesting subject and many detections have been successfully made with Parkes~\citep[]{2021MNRAS.507.3238Y,2021arXiv210604821T},  Molonglo Observatory Synthesis Telescope (MOST)~\citep[]{2019ATel13363....1G}, The Canadian Hydrogen Intensity Mapping Experiment (CHIME)~\citep[]{Masui2021First}, and other observatories~\citep[]{2019ATel13376....1S,2019ARep...63...39F,2020ApJ...896L..40P,2020nova.pres.6769K}. However, an important question that remains is "What is the origin of FRBs?". There have been numerous theories regarding their origins~\citep[]{2015IAUGA..2228939B,2016MPLA...3130013K,2021AAS...23732804M,2021ApJ...910L..18B} although none of those have been fully confirmed. Even though we do not understand their origins, FRBs can be used for many cosmological experiments - one of which is to test the famous Equivalence Principle of Einstein, which is the starting point of General Theory of Relativity~\citep[]{1908JRE.....4..411E}. The equivalence principle is a hypothesis which states - if some persons are free-falling, their experience of weightlessness will be the same as their experience in a space without gravity. This idea is referred to as Weak Equivalence principle (WEP). While the weak principle only focuses on the free-falling aspect of gravity~\citep[]{2006LRR.....9....3W,2009IAU...261.1001W,2014LRR....17....4W}, the "strong equivalence principle"~\citep[]{Bertotti_1990} is a more generalized statement. It states that the effect of acceleration and that of an equivalent gravitational field would always be the same. In this study, we would be focused on testing the WEP using the latest CHIME FRB data~\citep[]{Masui2021First}.

FRBs are generally classified into two types - repeating and non-repeating FRBs. The repeating signals are observed to repeat after certain intervals (the intervals might vary from hours to days and the repetitions are not necessarily periodic) while the non-repeaters are single pulses. The classification is simply observational and certain non-repeaters can turn out to be repeaters in the future. A necessary reason for the existence of two kinds of FRBs is probably associated with their origins according to studies by~\citet[]{2020MNRAS.494.2886H, 2021arXiv210604356P, 2020Natur.587...54C, 2021NatAs...5..414K, 2020Natur.587...59B, 2021AAS...23723603P}. These studies suggest that there is a possibility that the non-repeating FRBs are mostly originated from astronomical objects like Black Holes, White Dwarfs and Neutron Stars while repeating FRBs originated from supernova remnants, young pulsars and magnetars. Certain more interesting properties of FRBs are their emission frequencies which lie mostly in the 400 MHz to 1 GHz range. FRBs are presumed to have powerful engines, as the amount of energy release in a few milliseconds can be equivalent to the energy released by the Sun in a few days~\citep[]{2021arXiv210710113P}. The latest data released by CHIME is a catalogue consisting of 535 FRB sources in the frequency range of 400 MHz to 1 GHz, approximately.

The test of WEP is based on the "Shapiro Time Delay" method which is considered to be the fourth test of GR~\citep[]{1964PhRvL..13..789S}. The other three tests were defined by Einstein himself as follows - (1) The precision of perihelion of Mercury's orbit, (2) The gravitational redshift of light and (3) The light-deflection by the Sun as observed during eclipses~\citep[]{PhysRevLett.116.221101,doi:10.1146/annurev-astro-091918-104423,PhysRevD.100.104036}. In simple words, the Shapiro Time Delay test uses the idea that when a ray of light passes through a gravitating object, its path will be slightly deflected according to the geodesic or the curvature of the space. This leads to a small delay in its arrival in comparison to an unperturbed path if no gravitational field is present. The amount of curvature of space is defined in the Parameterized Post Newtonian (PPN) formalism using the gamma parameter ($\gamma$). $\gamma$ defines how much curvature is produced by a unit test mass~\citep[]{Misner:1973prb}. The test mass particle can be a massless particle like a photon, or the hypothetical graviton, or the neutrino. In our study, the test particles are the FRB photons~\citep[]{PhysRevLett.115.261101}. In the case of general relativity, $\gamma$=1. A slight deviation thus indicates that a photon is not following the space-time geodesic, violating the WEP. 

In this study, we considered many cosmological models to verify our results. All these models are fiducial $\Lambda$ Cold Dark Matter (LambdaCDM) model with the cosmological parameters ($\Omega_{\rm m}$,$\Omega_{\rm \Lambda}$,$\Omega_{\rm b}$,$H_{\rm 0}$). The model parameters in our study are Planck13 (0.307,\,0.693,\,0.0486,\,67.8) ~\citep[]{2014A&A...571A..16P}, Planck15 (0.307,\,0.693,\,0.0486,\,67.7) ~\citep[]{2016A&A...594A..13P}, and WMAP9 (0.287,\,0.713,\,0.02256,\,69.3) ~\citep[]{2013ApJS..208...19H}.

The paper is arranged in the following way. We discuss the theoretical considerations behind our analysis in Section \ref{section:Theoretical Model}. We discuss the FRB dataset that we have used in Section \ref{section:Dataset}. In Section \ref{section:Results}, the results related to the study can be found, followed by a discussion comparing to previous studies in Section \ref{section:Discussion}. The conclusions are in Section \ref{section:Conclusion}.

\section{Theoretical Model}
\label{section:Theoretical Model}


For any transient beam observed from a cosmic distance with $z > 1$, like FRBs, we are required to consider the following reasons for time-delay. The time-delay can be expressed in terms of the following equation~\citep[]{WANG2020100571} under the observation by a telescope:
\begin{align}
\label{eqn:1}
    \Delta t_{\rm obs} = \Delta t_{\rm DM} + \Delta t_{\rm ini} + \Delta t_{\rm spe} + \Delta t_{\rm LIV} + \Delta t_{\rm gra},   
\end{align}
where $\Delta t_{\rm obs}$ denotes the total possible time-delay of a particular FRB transient between different frequencies. $\Delta t_{\rm DM}$ is the time-delay between different frequencies of light calculated from the dispersion due to the ionised medium between the source and the observer. This dispersion measure (DM) may be caused by the interaction of photons with the host galactic medium, the intergalactic medium, and the Milky Way. $\Delta t_{\rm ini}$ is the initial time-delay of emission between two photons which is an intrinsic property of the source. $\Delta t_{\rm spe}$ is the time-delay due to different time-dilation of  photons with different non-zero rest mass which is a rather negligible special relativistic effect. $\Delta t_{\rm LIV}$ is the time-delay due to Lorentz Invariance Violation. $\Delta t_{\rm gra}$ is the Shapiro Time-Delay~\citep[]{PhysRevLett.13.789} between two photons with different energies passing by a gravitational field V(r):
\begin{align}
\label{eqn:2}
\Delta t_{\rm gra} = \frac{\gamma_{\rm 2} - \gamma_{\rm 1}}{c^3} \int_{r_{\rm 0}}^{r_{\rm e}} V(r) \,dx,
\end{align}
where $c$ is the speed of light, $r_{\rm o}$ and $r_{\rm e}$ are positions of source and observer, and $\gamma_{\rm 1}$ and $\gamma_{\rm 2}$ are parameterized post-Newtonian formalism of two photons of an FRB. In general, for all metric theories, WEP remains valid i.e. any test particle independent of energy should follow the space-time geodesic and hence $\gamma_{\rm 2}$ = $\gamma_{\rm 1}$ = 1. Now, in Eq. \ref{eqn:1}, according to the approximation of  ~\citet[]{WANG2020100571,2020PDU....2900571W}, the  terms $\Delta t_{\rm spe}$ and $\Delta t_{\rm LIV}$ are negligible in comparison to the other time-delays, and $\Delta t_{\rm ini}$ > $0$ is considered. Thus, Eq. \ref{eqn:1} changes to the following inequality:
\begin{align}
\label{eqn:3}
\Delta t_{obs} - \Delta t_{DM} > \Delta t_{gra}    
\end{align}
This inequality explains that by subtracting the effect of dispersion measure (DM) from the total time delay, we can clearly observe the time-delay caused by the Shapiro effect. In many of the earlier papers like~\citet[]{Nusser_2016}, the Shapiro time delay is calculated based on local universe models like the Large Scale Structure models (like the Laniakea Super-cluster) or considering a host galaxy as the reasons. However, for huge cosmological distances where $z\gtrsim1$, it is not the proper calculation. In fact, for cosmological sources, $t_{gra}$ and $\Delta t_{gra}$ do not have a monotonic increment, as shown in ~\citet[]{Minazzoli_2019}. Thus, to calculate $\Delta t_{gra}$, we have to consider all the gravitational sources in the path of the radiation. This is rather difficult and impractical, considering the fact that we do not have enough knowledge and the calculations become more complicated at $z\gtrsim1$. Therefore, we consider an analytic model-based solution where we consider the average matter distribution. In this way, the average Shapiro time delay $t_{gra,avg}$ can be expressed as a summation of the following two terms: 
\begin{align}
\label{eqn:4}
    t_{\rm gra,avg} = t_{\rm \Lambda} + t_{\rm matter},
\end{align}
where,
\begin{align}
\label{eqn:5}
    t_{\rm \Lambda} = \frac{\Omega_{\rm \Lambda}H^2_{\rm 0}}{12c^3}d^3_{\rm s},
\end{align}
and
\begin{align}
\label{eqn:6}
    t_{matter} = \frac{\Omega_{\rm m}H^2_{\rm 0}}{6c^3}d^3_{\rm s},
\end{align}
where $d_{\rm s}$ is the co-moving distance of the cosmological source. The co-moving distance is defined as the distance between any two cosmological objects after factoring out the expansion of the universe, so as to take note of any other variation that can lead to a change of distance. Eq. \ref{eqn:6} is consistent~\citep[]{2019PhRvD.100j4047M} up to at least $\sim$ 400 Mpc ($z$ $\sim$ 0.1) when Shapiro Time Delay is calculated from observed galaxy clusters.

Now, in this paper, we focus on a time lag under gravity between two photons with different energies. Because, we focus on the time-lag due to the gravitational effect, we assume that the $t_{\rm \Lambda}$ term is cancelled  out between different frequencies while calculating $\Delta t_{\rm gra}$.The time delay for the first and second photon can thus be expressed as -
\begin{align}
\label{eqn:7}
    t_{\rm matter,\gamma_{1}} = \gamma_{\rm 1}\frac{\Omega_{\rm m}H^2_{\rm 0}}{6c^3}d^3_{\rm s}
\end{align}
and
\begin{align}
\label{eqn:8}
    t_{\rm matter,\gamma_{2}} = \gamma_{\rm 2}\frac{\Omega_{\rm m}H^2_{\rm 0}}{6c^3}d^3_{\rm s}
\end{align}
respectively. Thus, we express $\Delta t_{\rm gra}$ in terms of $t_{\rm matter}$ and the PPN parameters in the following form:
\begin{align}
\label{eqn:9}
\Delta t_{\rm gra} = t_{\rm matter,\gamma_{2}} - t_{\rm matter,\gamma_{1}}   = (\gamma_{\rm 2} - \gamma_{\rm 1}) \frac{\Omega_{\rm m}H^2_{\rm 0}}{6 c^3} d^3_{\rm s}, 
\end{align}
where, $\gamma_{\rm 1}$ and $\gamma_{\rm 2}$ are the PPN parameter values of photon-1 and photon-2 respectively. From Eq. \ref{eqn:3} and \ref{eqn:9}, we obtain,
\begin{align}
\label{eqn:10}
    \Delta \gamma := \gamma_{\rm 2} - \gamma_{\rm 1} < (\Delta t_{\rm obs} - \Delta t_{\rm DM}) \frac{6 c^3}{\Omega_{\rm m}H^2_{\rm 0}d^3_{\rm s}},
\end{align}
In the case of FRB data, the deviation of WEP can be observed in the uncertainties of the observed Dispersion Measure ($\rm DM_{\rm obs}$) i.e. $\rm \delta \rm DM_{\rm obs}$. Therefore, evidence might exist where the value of observed frequency ($\nu^{\rm -2}_{\rm obs}$) should effect the violation of WEP. It is because $t_{\rm obs}$ can be represented by $\Delta t_{\rm DM}$ and also $\nu^{\rm -2}_{\rm obs}$. If both the violation of WEP and $\Delta t_{\rm DM}$ show an equivalent dependence on $\nu^{\rm -2}_{\rm obs}$, the two effects are not distinguishable~\citep[]{Ioka_2003,Macquart_2020,Inoue_2004}. Until now, current studies of FRBs or any other sources have not found any such violation  due to frequency~\citep[]{1908JRE.....4..411E, Tingay_2016,Nusser_2016, Minazzoli_2019, Xing_2019}. 

The time lag due to the observed dispersion measure can be expressed using the following inequality as
\begin{align}
\label{eqn:11}
    \Delta t_{\rm DM} \simeq 4.15\,(\frac{\nu_{\rm obs}}{1\, GHz})^{\rm -2} \frac{DM_{\rm obs}}{10^{3} \rm pc\, cm^{-3}}\, \rm s,
\end{align}
 The uncertainty of $\Delta t_{DM}$ ($\delta \Delta t_{DM}$) is proportional to $\delta DM_{obs}$ as -
 \begin{align}
 \label{eqn:12}
     \delta \Delta t_{\rm DM} \simeq 4.15\,(\frac{\nu_{\rm obs}}{1\,GHz})^{-2} \frac{\delta DM_{\rm obs}}{10^{\rm 3} \rm pc\, \rm cm^{\rm -3}}\, \rm s,
 \end{align}
 The limit of $\delta \Delta t_{\rm DM}$ defines an upper limit constraint on WEP violation. Thus, $\delta DM_{\rm obs}$ puts an upper limit on $\Delta \gamma$. Here, we have considered $\delta \Delta t_{\rm DM}$ as the upper limit of $\Delta t_{\rm obs}$ - $\Delta t_{\rm DM}$ as in eqn \ref{eqn:10}:
 \begin{align}
 \label{eqn:13}
     \Delta t_{\rm obs} - \Delta t_{\rm DM} < \delta \Delta t_{\rm DM},
 \end{align}

\section{Data}
\label{section:Dataset}
We adopt the latest FRB catalogue released by Canadian Hydrogen Intensity Mapping Experiment (CHIME) as of April, 2021~\citep[]{Masui2021First}. 
Here, we use classification in the catalogue to separate the repeating and non-repeating cases.
However, see Chen et al. (2021 submitted) for a more efficient classification using Machine Learning. We use redshift and its uncertainty of each FRB  calculated from the observed dispersion measure (DM$_{\rm obs}$) by Hashimoto et al. (2021 in prep.). Here, we consider the redshift, redshift uncertainty, $DM_{\rm obs}$, $\delta DM_{\rm obs}$, $\nu_{\rm obs}$, and observed bandwidth for our analysis. The observed frequency $\nu_{\rm obs}$ is the peak-frequency for each FRB sub-burst in our study. The peak-frequency is the frequency value at which the amplitude of the signal is highest. 

\begin{figure}
   \centering
    \includegraphics[width=\columnwidth]{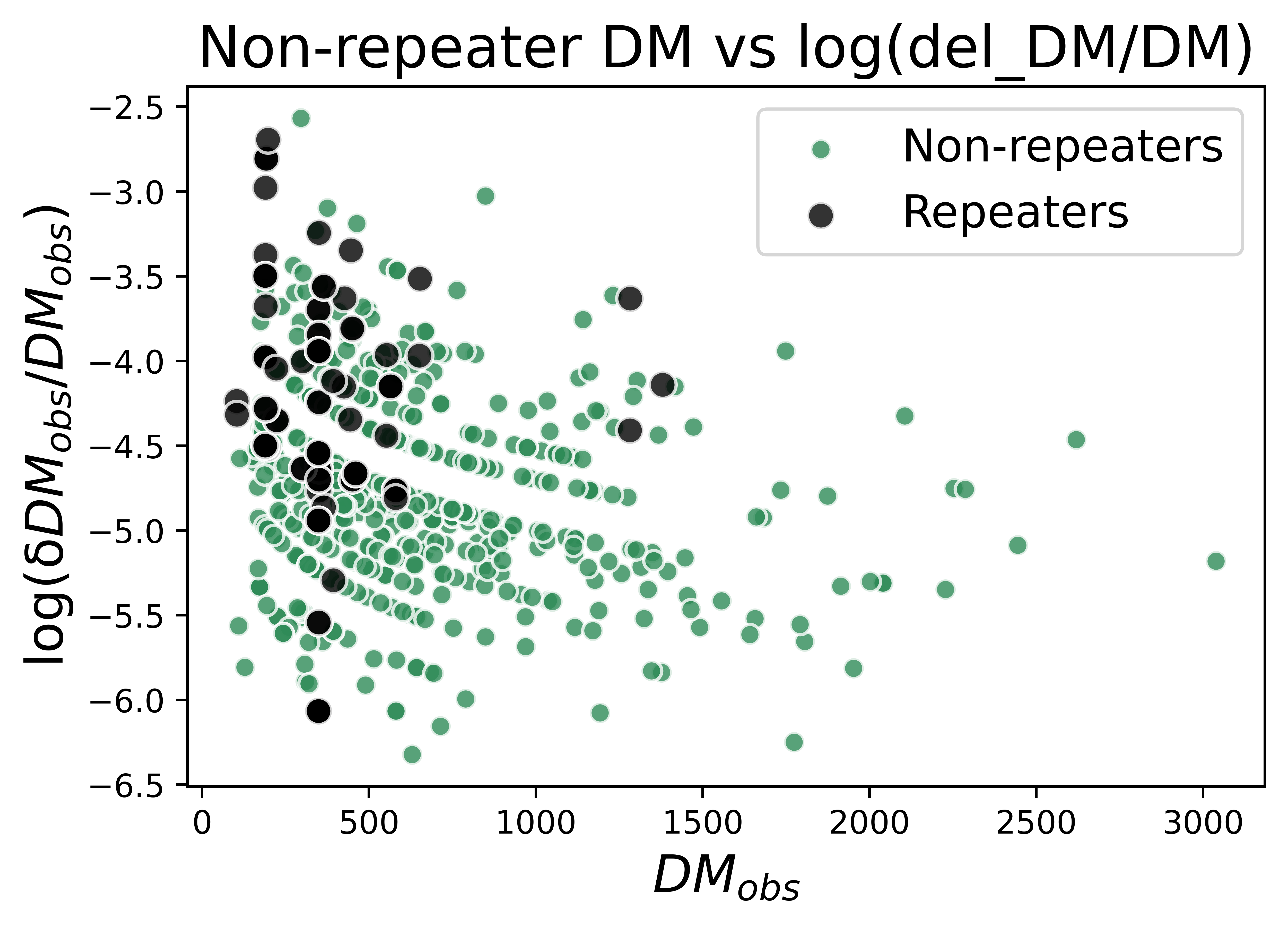}
    \caption{log($\delta DM_{\rm obs}$/$DM_{\rm obs}$) vs $DM_{\rm obs}$ for the case of Repeaters and Non-repeaters using the CHIME dataset. In the case of repeaters, $DM_{\rm obs}$ and $\delta DM_{\rm obs}$ are calculated for each repeater having multiple values obtained at periodic intervals}
    \label{fig:1}
\end{figure}

In the Fig \ref{fig:1}, we show the ratio of ($\delta DM_{\rm obs}$/$DM_{\rm obs}$) as a function of $DM_{\rm obs}$, for non-repeating and repeating FRBs. For the non-repeating FRBs, log($\delta DM_{\rm obs}$/$DM_{\rm obs}$) $\sim -6.3$ while in the case of repeating FRBs, the value is around $-6$. Non-repeaters show slightly lower values because their brightnesses are higher compared to the repeaters. On average, therefore, $\delta$DM$_{\rm obs}$ is slightly smaller for non-repeating FRBs with higher signal-to-noise ratios. In comparison to an earlier study by Hashimoto et. al. (submitted) using earlier FRBCAT data where the non-repeaters lower limit of log($\delta DM_{\rm obs}$/$DM_{\rm obs}$) was around -5, this result is around one order of magnitude lower, which implies the CHIME dataset provides a more accurate $DM_{\rm obs}$. Thus, we can provide a tighter constraint on the time lag between different photon energies (having different energies and frequencies). 

\section{Results}
\label{section:Results}
\begin{figure}

    \centering
    \includegraphics[width=\columnwidth]{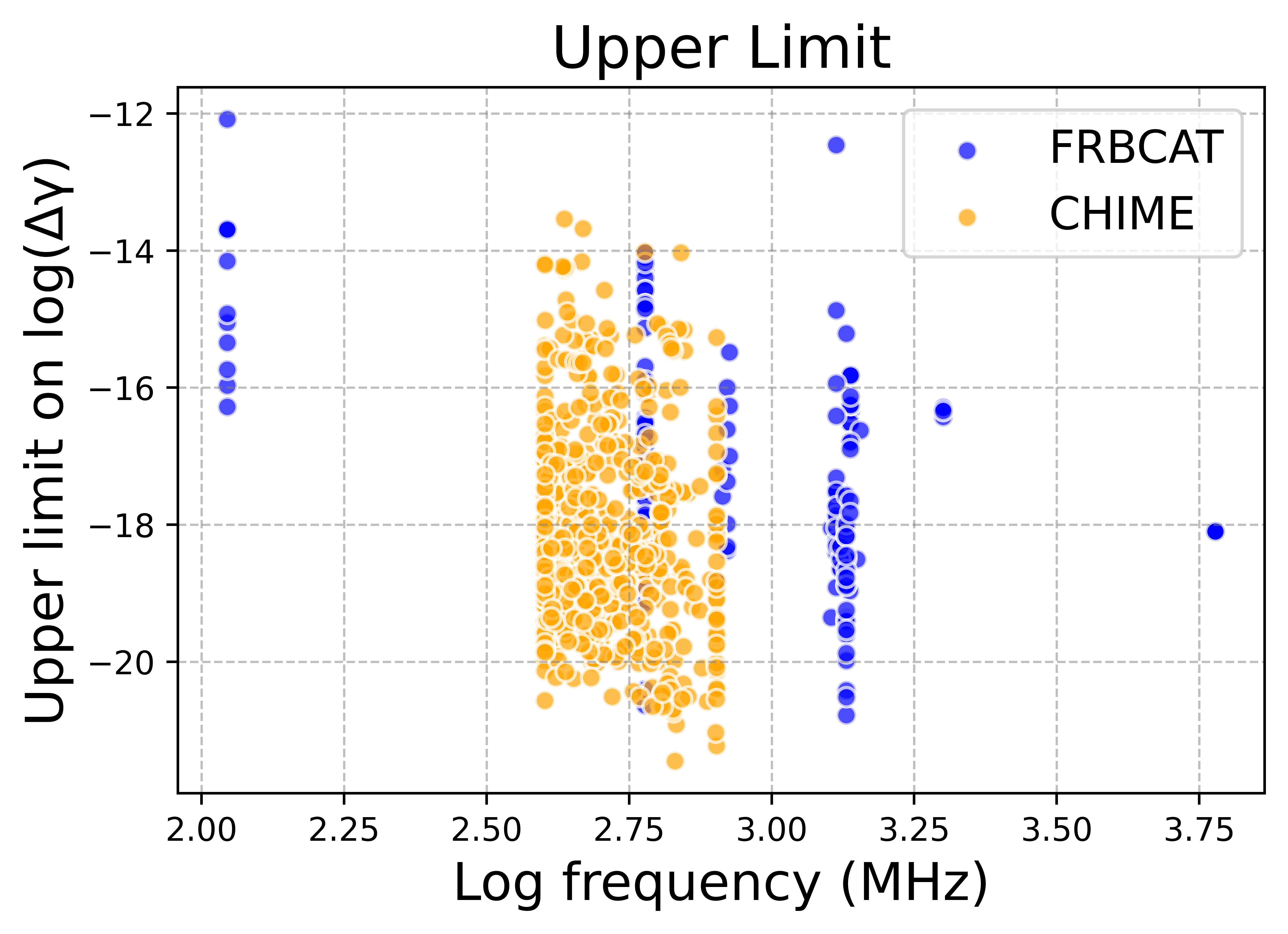}
    \includegraphics[width=\columnwidth]{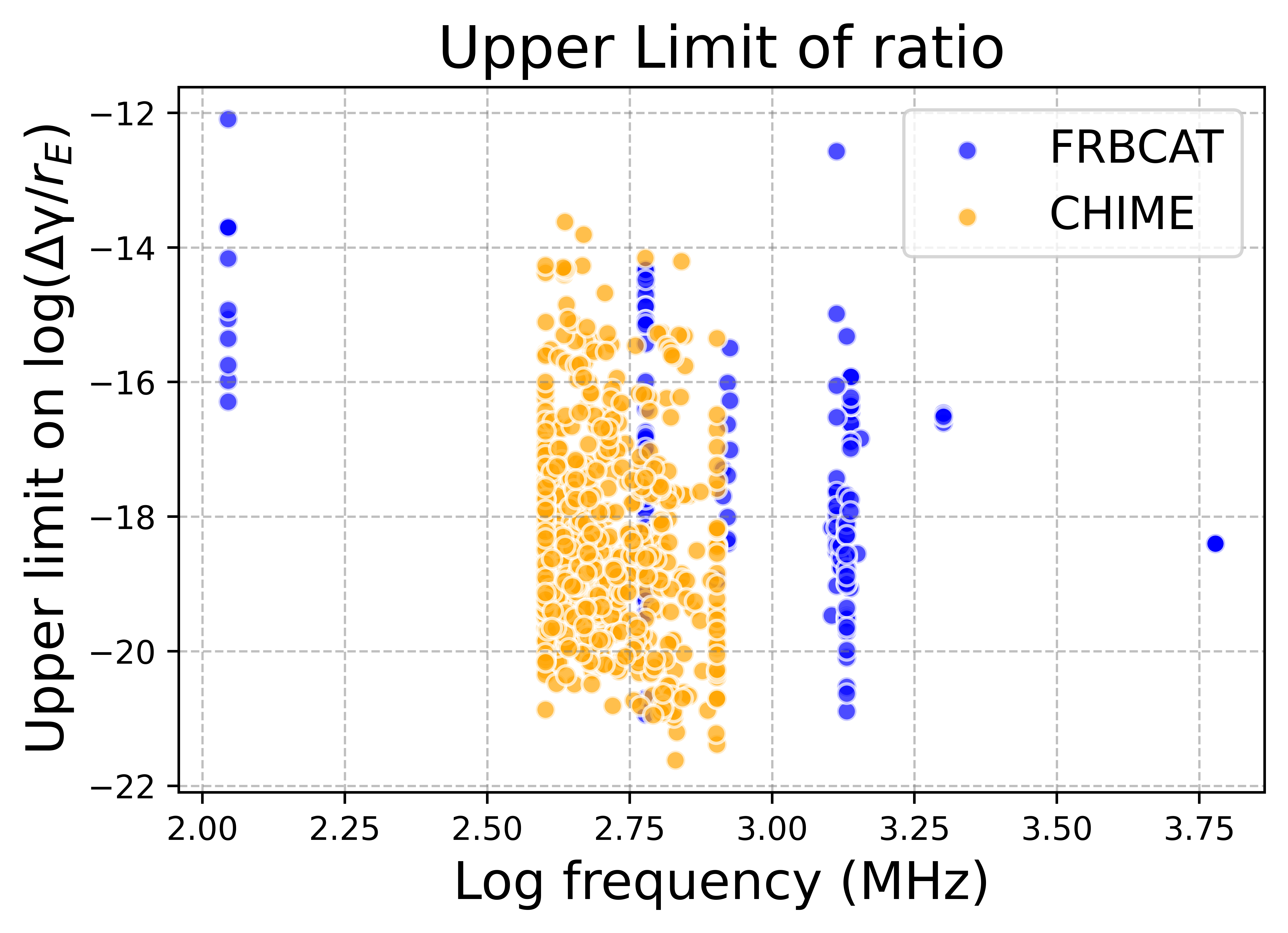}
    \caption{(a)Comparison of log($\Delta \gamma$) as a function of observed frequency between FRBCAT and CHIME data. The CHIME data clearly shows an increment in the order of magnitude of the log($\Delta \gamma$) value. The other FRB data mentioned in Section \ref{section:Discussion} are included in the FRBCAT data (b)Comparison of log($\Delta \gamma/r_{\rm E}$) as a function of observed frequency between FRBCAT and CHIME data. Here $r_{\rm E}$ is the ratio of energy i.e.$r_{\rm E}$ = $E_{\rm high}/E_{\rm low}$.The CHIME data clearly shows an increment in the order of magnitude of the log($\Delta \gamma$) value. The other FRB data mentioned in Section \ref{section:Discussion} are included in the FRBCAT data.}
    \label{fig:3}
\end{figure}

We aim to find the constraint on $\Delta \gamma$. According to our understanding from various previous studies~\citep[]{Tingay_2016,Nusser_2016, Xing_2019,PhysRevD.104.084025,Li_2021,Yang_2020,tangmatitham2019testing}, we can only consider the upper limit of $\Delta \gamma$ value to mark a constraint which essentially indicates an upper limit of the violation of WEP.  
\begin{figure}

    \centering
    \includegraphics[width=\columnwidth]{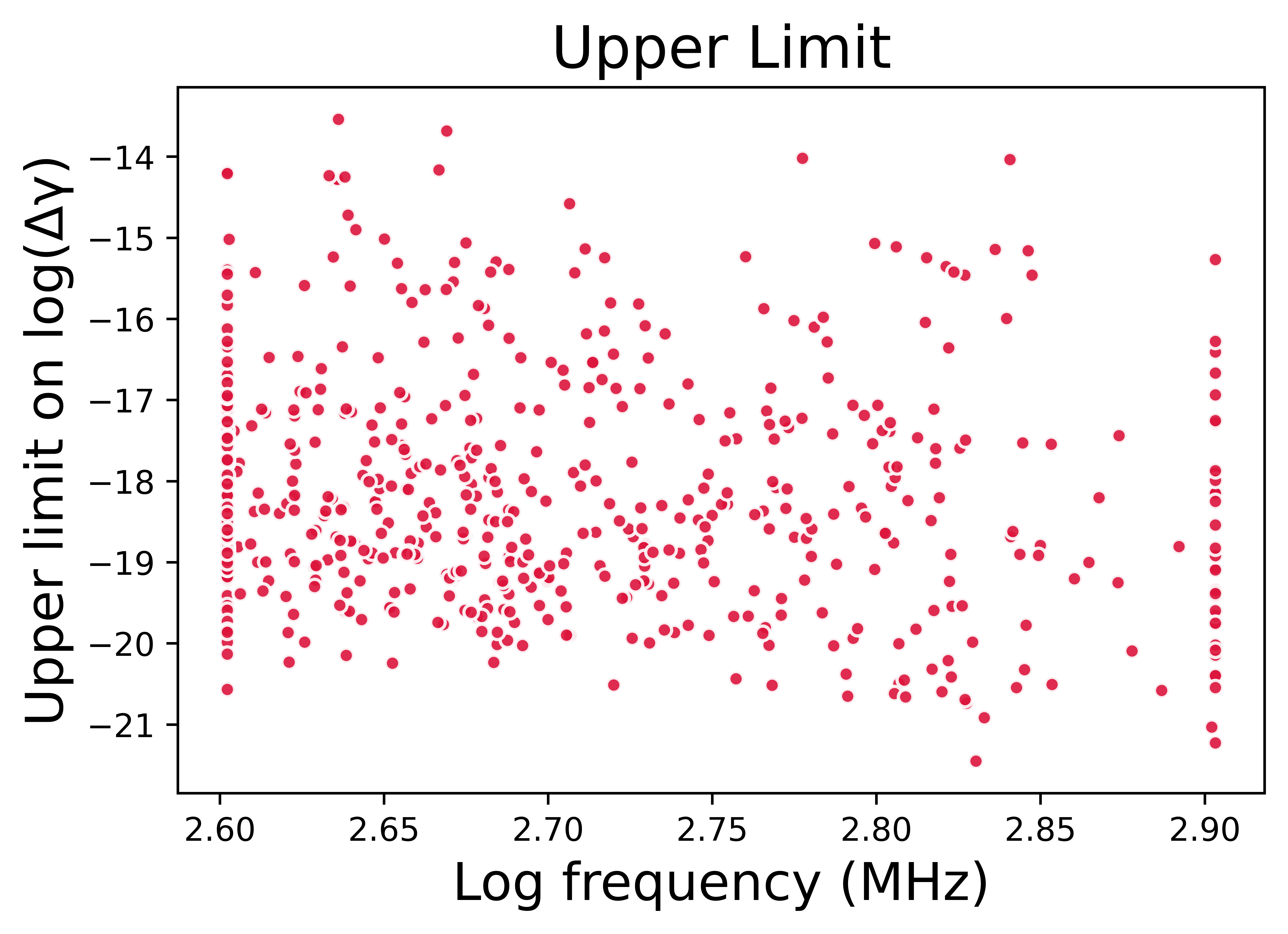}
    \includegraphics[width=\columnwidth]{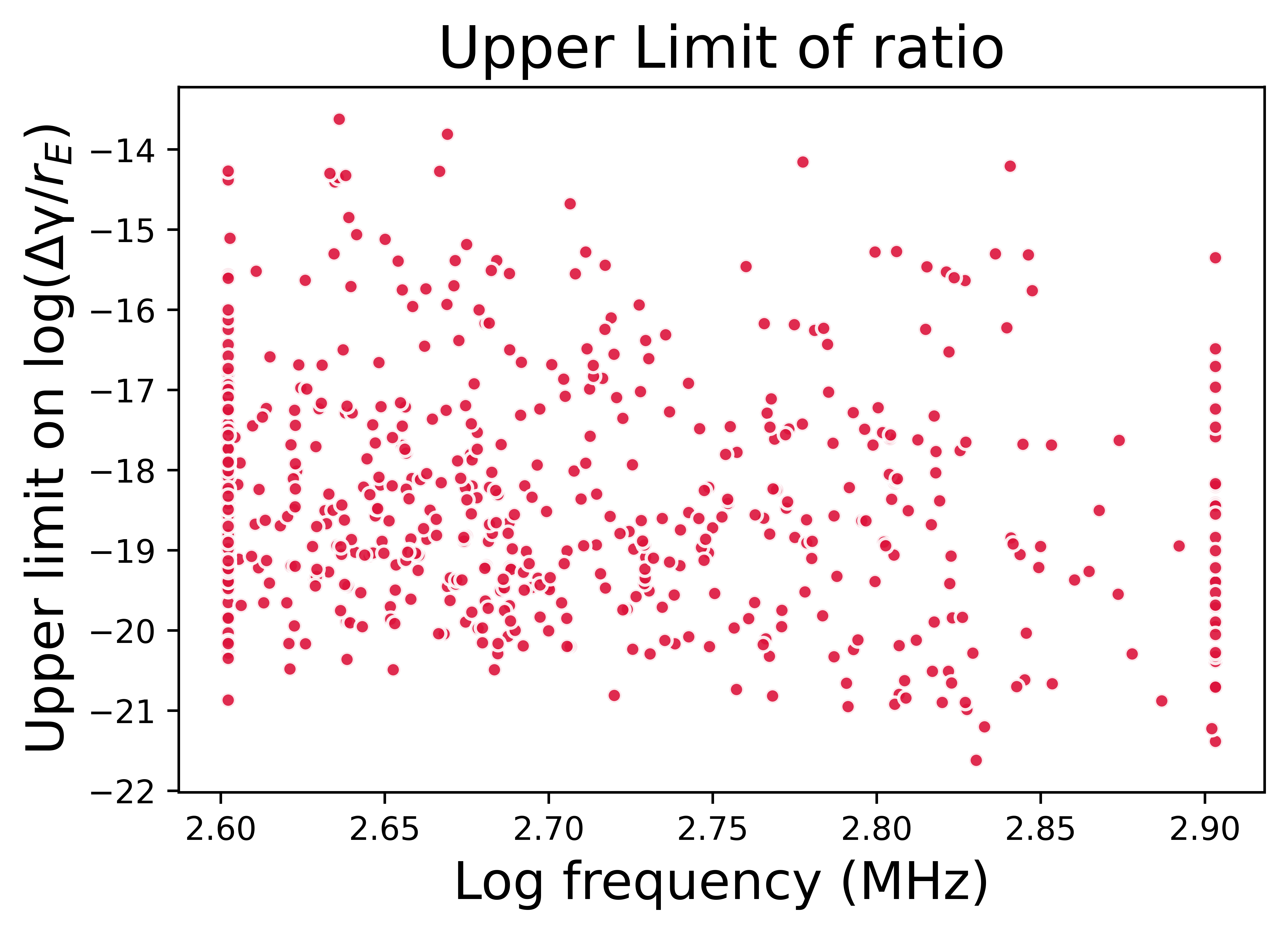}
    \caption{(a)log($\Delta \gamma$) as a function of observed frequency, (b)log($\Delta \gamma/r_{E}$) as a function of observed frequency. Here $r_{E}$ is the ratio of energy i.e.$r_{E}$ = $E_{high}/E_{low}$ }
    \label{fig:2}
\end{figure}
According to our calculation from Eqs. \ref{eqn:10},\ref{eqn:12} and \ref{eqn:13} we find a limit on $\Delta \gamma$ as a function of the observed frequency. In Fig \ref{fig:2}, the lowest upper limit of $\Delta \gamma$ of the distribution is approximately around  -21 in the logarithmic scale. The most stringent constraint in our work is shown by FRB20181117C as $\Delta \gamma$ = $-21.58 ^{+0.10}_{-0.12}$. For FRB20181117C, the value of $z$ = $1.83 \pm  0.28$, $DM_{\rm obs}$ = $1773.739$ \rm pc $\rm cm^{-3}$, $\delta DM_{obs}$ = $0.001$ $\rm pc$ $\rm cm^{-1}$ and $\nu_{\rm obs}$ = $676.5 \, \rm MHz$. Interestingly, this constraint is so far the best possible constraint we have been able to obtain using FRB data. The most recent constraint as described by Hashimoto et. al. using the FRBCAT data~\citep[]{2016PASA...33...45P, Hashimoto_2020} also showed rather promising results for FRBs as they have shown a limit of $-20$ which is a tighter constraint than that from neutrinos, gamma-ray bursts (GRBs), and gravitational waves (GWs). A comparison between the FRBCAT data~\citep[]{2016PASA...33...45P, Hashimoto_2020} and the current CHIME FRB data~\citep[]{Masui2021First} is shown in Fig. \ref{fig:3}. Both these results give us an insight into the FRBs that they can act an effective probe to test the WEP. 

The average energy difference between the photons is around $\sim$ $20\%$~\citep[]{1908JRE.....4..411E, Tingay_2016,Nusser_2016, Xing_2019}. This is comparatively a lower energy variation with respect to that obtained from strong GRBs and X-ray Pulsar radiation. Hence, the deviation of WEP becomes more prominent due to high energy difference if the value of $\gamma$ depends on energy. In that case, we might not obtain a resolute evidence regarding whether WEP is violated or not. Therefore,~\citet[]{Tingay_2016} has introduced a new representation where log($\Delta \gamma$/$r_{\rm E}$) should be a function of the observed frequency ($\nu_{\rm obs}$). Here, $r_{E}$ is the ratio of the higher and lower energy of a signal. Thus, $r_{\rm E}$ $:=$ $E_{\rm high}/E_{\rm low}$ = $\nu_{\rm high}/\nu_{\rm low}$. From our sample we obtain that for FRB20181117C, the value of log($\Delta \gamma/r_{E}$) = $-21.75 ^{+0.10}_{-0.14}$ gives a stringent constraint. This result is about one order of magnitude lower than the previous result obtained from FRBCAT data developed by~\citet[]{2016PASA...33...45P, Hashimoto_2020}. We have considered a number of cosmological models as given in Section \ref{section:Introduction} to check any variation in our results. Through analysis, we find that the variations are negligible in terms of the cosmology parameters.

\section{Discussion}
\label{section:Discussion}
Constraining the PPN parameter $\Delta \gamma$ using FRBs has been done in many previous studies like~\citet[]{Tingay_2016,Nusser_2016, Xing_2019}. According to~\citet[]{PhysRevLett.115.261101}, considering FRB 110220 having frequencies between 1.18 to 1.52 GHz, they have obtained log $\Delta \gamma$ < -7.6. In their formulation, $\Delta t_{obs}$ is primarily dominated by WEP violation rather than dispersion measure. However, according to our formulation, for same FRB110220, the log $\Delta \gamma$ < -19.87. ~\citet[]{Tingay_2016} used FRB150418 in the frequency range 1.2 to 1.5 GHz, yielding a lower limit of log $\Delta \gamma$ < -7.7. They also calculated the limit to be lower up to -9 based on the assumption that the WEP is $~5\%$ masked by the uncertainty of DM$_{obs}$. Using the same FRB data, ~\citet[]{Nusser_2016} suggested using the gravitational field of large-scale structures like the Laniakea Supercluster~\citet[]{Tully_2014} rather than the Milky Way to measure the WEP constraint. They have obtained a lower limit of log $\Delta \gamma$ < -12. In our calculation, the limit of log $\Delta \gamma$ for FRB150418 is < -18.05. Using the ~\citet[]{Nusser_2016} model as described above, \citet[]{Xing_2019} studied the repeated burst of FRB121102 in the frequency range of 1.344 to 1.374 GHz. The time lag of their sub-burst samples were around $\Delta t_{obs}$ = 0.4 ms, which provides log $\Delta \gamma$ < -15.6. However, using the similar sample for FRB121102, our formulations have obtained a log $\Delta \gamma$ < -16.5. Again, an extended sub-burst set of FRB121102 in the frequency range of 4 to 6 GHz yields a limit of log $\Delta \gamma$ < -18.39. All the previous measurements have used the approximation of the Minkowski metric with linear perturbation. However, ~\citet[]{Minazzoli_2019} reconsidered the fact and concluded that such an assumption is not applicable for cosmological sources like FRBs. All the previous measurements were based on $z < 1$, whereas we have tried to consider FRB sources to be originated from cosmological distances where $z \gtrsim 1$.
Our approach follows the argument where the exact value of $\delta DM_{\rm obs}$ is considered instead of assuming the 5\% typical uncertainty of DM$_{\rm obs}$ is assumed ~\citep[]{Tingay_2016}. However, we have used the latest FRB data obtained by CHIME as of April 2021. In our study, we have obtained the lowest possible value of both log $\Delta \gamma$ and log ($\Delta \gamma/r_{\rm E}$) cases, considering same particles with different energies.

\section{Conclusion}
\label{section:Conclusion}
Fast Radio Bursts (FRBs) are radio bursts coming from yet unknown sources at cosmological distances. The time-lag between different energies of FRB signals depends on the dispersion measure. However, they are also hypothesized to be dependent on the frequency following the inverse square law ($\nu^{-2}_{obs}$). Thus, we have tested the Weak Equivalence Principle (WEP) using FRB data. WEP is the hypothetical time-lag between photons with different frequencies caused by a gravitational field. Hence, the violation of WEP, if it exists, can be detected in the observational uncertainties of the Dispersion Measures of FRBs. Note that it is our assumption that the frequency dependence of WEP violation does not mimic that of dispersion.

We have used the concept of "Shapiro Time Delay" using the post-Newtonian parameter $\gamma$ which explains how much gravitational curvature is caused by unit mass. Measuring the variation $\Delta \gamma$ as a function of observed frequency, we have derived the strictest constraint of log($\Delta \gamma$) at $-21.58^{+0.10}_{-0.12}$ and log($\Delta \gamma/r_{\rm E}$) at $-21.75 ^{+0.10}_{-0.14}$ for FRB20181117C whose z = $1.83 \pm 0.28$ , $DM_{obs}$ = $1773.739$ $\rm pc$ $cm^{\rm -3}$, $\delta DM_{\rm obs}$ = $0.001$ $\rm pc$ $cm^{\rm -1}$ and $\nu_{\rm obs}$ = $676.5$ $\rm MHz$. Our new limit is about one order of magnitude tighter than the previous constraint obtained from the FRBCAT data developed by Hashimoto et. al. (submitted) and about five orders of magnitude tighter than any other previous studies.

\section*{Acknowledgements}
We thank the anonymous referee for careful reading of the manuscript and many insightful comments.
KS was supported by the Centre for Informatics and Computation in Astronomy (CICA) at National Tsing Hua University (NTHU) through a grant from the Ministry of Education of Taiwan. KS, TH, TG and team acknowledge the supports by the Ministry of Science and Technology of Taiwan through grants 108-2628-M-007-004-MY3 and 110-2112-M-005-013-MY3, respectively. This research has made use of NASA's Astrophysics Data System.

\section*{Data Availability}

The data underlying this article are available at \url{https://www.chime-frb.ca/catalog}.
The redshifts and their uncertainties of FRBs will be available at the article online supplementary material after the publication of Hashimoto et al. 2021 in prep.



\bibliographystyle{mnras}
\bibliography{WEP_paper} 








\bsp	
\label{lastpage}



\end{document}